\documentclass[twocolumn, superscriptaddress, prb, preprintnumbers, showpacs]{revtex4-1}

\usepackage{ifthen} 
\usepackage{graphicx}
\usepackage{amsmath}

\def\bra#1{\langle#1\vert}
\def\ket#1{\vert#1\rangle}
\def\braket#1#2{\langle#1\vert#2\rangle}
\def\exv#1#2#3{\langle#1\vert#2\vert#3\rangle}
\def\bfk{{\mathbf{k}}}

\begin{document}

\title{Projector self-consistent DFT+$U$ using non-orthogonal generalized
  Wannier functions}

\author{David D. O'Regan} 
\email{ddo20@cam.ac.uk}
\affiliation{Cavendish Laboratory, University of Cambridge,
  J. J. Thomson Avenue, Cambridge CB3 0HE, United Kingdom}

\author{Nicholas D. M. Hine}
\affiliation{Cavendish Laboratory, University of Cambridge,
  J. J. Thomson Avenue, Cambridge CB3 0HE, United Kingdom}
\affiliation{The Thomas Young Centre, 
Imperial College London, London SW7 2AZ, United Kingdom}

\author{  Mike C. Payne}
\affiliation{Cavendish Laboratory, University of Cambridge,
  J. J. Thomson Avenue, Cambridge CB3 0HE, United Kingdom}
  
\author{Arash A. Mostofi}
\affiliation{The Thomas Young Centre, 
Imperial College London, London SW7 2AZ, United Kingdom}

\date{\today{}}

\begin{abstract}
We present a formulation of the density-functional theory + Hubbard
model (DFT+$U$) method that is self-consistent over the choice of
Hubbard projectors used to define the correlated subspaces. 
In order to overcome the arbitrariness in this choice, we propose
the use of non-orthogonal generalized Wannier functions (NGWFs) as
projectors for the DFT+$U$ correction. 
We iteratively refine these NGWF projectors and, hence, the 
DFT+$U$ functional, such that the correlated subspaces are
fully self-consistent with the DFT+$U$ ground-state.
We discuss the convergence characteristics of this algorithm and
compare ground-state properties thus computed with those calculated
using hydrogenic projectors.
Our approach is implemented within, but not restricted to, a
linear-scaling DFT framework, opening the path to DFT+$U$ 
calculations on systems of unprecedented size.

\end{abstract}

\pacs{71.15.Mb, 31.15.E-, 71.15.Ap    \quad
(Accepted for Physical Review B Rapid Communications)}

\maketitle

The physics of localized electrons bound to transition metal or
Lanthanoid ions is important for understanding and harnessing the
behaviour of complex systems such as
molecular magnets~\cite{angewchem.33.385}
inorganic catalysts~\cite{nature.414.345}
and the organometallic molecules that facilitate some of the most
critical chemical reactions in biochemistry~\cite{chemrev.96.2239}.

Despite its success at predicting ground-state properties of
materials, Kohn-Sham density-functional 
theory (DFT)~\cite{PhysRev.136.B864,
*PhysRev.140.A1133} fails to describe the
physics of such ``correlated-electron'' systems
   when local or
semi-local exchange-correlation (XC) functionals are used, often
predicting results that are not only quantitatively but qualitatively 
inconsistent with experiment~\cite{PhysRevB.30.4734,
*PhysRevB.57.1505}.  The origin of
this apparent failure has been understood since the work of Perdew
\emph{et al.}~\cite{PhysRevLett.49.1691} and is related 
to the unphysical curvature of the energy functional with respect to
electronic occupation number~\cite{PhysRevB.71.035105,
*PhysRevLett.97.103001,
AronJ.Cohen08082008} inherent to such functionals unless 
a self-interaction correction is 
employed~\cite{PhysRevLett.65.1148}.

DFT + Hubbard $U$
(DFT+$U$)~\cite{PhysRevB.44.943, *PhysRevB.48.16929}
 is a simple, computationally 
inexpensive method for improving the description of on-site Coulomb
interactions provided by conventional XC functionals
and, hence, for extending the range of applicability of DFT
 to strongly-correlated materials.

The principle of DFT+$U$ is to divide the system into a delocalized,
free electron-like part, the ``bath'', which is well-described by
conventional XC functionals, and a set of ``correlated sites'' which
is not. The XC functional for electrons associated with these sites is
then explicitly augmented with screened Coulomb interactions, the form 
of which are inspired by the Hubbard-model~\cite{hubbard1,
  *hubbard2,*hubbard3},
together with a double-counting term to correct for the component
already included within the XC functional.

The correlated sites are defined by a set of $3d$ and/or $4f$
atomic-like orbitals, or ``Hubbard projectors'', that are chosen
\emph{a priori}. Projector functions that are commonly used include
hydrogenic wavefunctions~\cite{PhysRevB.71.035105,*PhysRevLett.97.103001},
maximally-localized Wannier functions~\cite{PhysRevB.77.085122},
and LMTO-type orbitals~\cite{PhysRevB.44.943,PhysRevB.57.1505}. This
arbitrariness constitutes an unsatisfactory, adjustable parameter in the
DFT+$U$ method. 

In this article, we present an approach in which the
ambiguity in the choice of Hubbard projectors is removed, and in which
they are determined self-consistently with respect to the DFT+$U$
ground-state. We first outline the theoretical framework of our 
approach, and present results of calculations on ligated
iron porphyrin. We examine the adequacy of hydrogenic orbitals
as Hubbard projectors and, in particular, the sensitivity of the
results to the form of these orbitals. We show that optimized
non-orthogonal generalized Wannier functions (NGWFs) provide an
unambiguous and natural choice for Hubbard projectors and we introduce
a technique for self-consistently delineating the subspaces in which
correlation effects play an important role.

Our implementation is within the framework of linear-scaling DFT,
however, the same self-consistent projector methodology may be applied
to any DFT approach that solves for localized Wannier-like 
functions (either directly, or indirectly in a post-processing step
using an interface to a code such as
Wannier90~\cite{wannier90}). Furthermore, our approach may be readily
combined
with recently-proposed methods to calculate $U$
parameters from
first-principles~\cite{PhysRevB.71.035105,*PhysRevLett.97.103001,
PhysRevB.58.1201}, facilitating 
entirely parameter-free and self-consistent DFT+$U$ calculations.

The Hubbard energy correction term in DFT+$U$ can be interpreted as a
functional that penalizes the unphysical non-integer occupancy of the
spatially localized $d-$ or $f-$orbitals, those that
are most prone to the 
spurious self-interaction present in standard DFT XC functionals. 

We use a rotationally-invariant correction term, 
\begin{equation}
E_U = \sum_{I \sigma} \frac{U^{(I)(\sigma)}}{2} \left[ \sum_{m}
 n_{m}^{\; m} - \sum_{m m'} n_{m}^{\; m'}
  n_{m'}^{\; m}\right]^{(I)(\sigma)},
\label{eq:EU}
\end{equation}
where $U^{(I)(\sigma)}$ represents the screened Coulomb
repulsion between electrons of spin $\sigma$, localized
on the correlated site $I$. Eq.~(\ref{eq:EU}) is, 
in effect, a penalty functional for
deviation from idempotency of the projection of the single-particle
density-matrix onto each correlated subspace.

The occupancy matrix 
in the case of a set of $M$ \emph{non-orthogonal} Hubbard projectors
$\ket{\varphi^{(I)}_{m}}$, $m\in\{1,\ldots,M\}$, localized on site
$I$, is given by  
\begin{equation}
n^{(I)(\sigma) m'}_{ m} = \sum_{i \bfk} f_{i \bfk}^{(\sigma)} 
\exv{\psi_{i \bfk}^{(\sigma)}}{\hat{P}^{(I)m'}_{m}}{\psi_{i
    \bfk}^{(\sigma)}},
\end{equation}
where $\psi_{i\bfk}^{(\sigma)}$ is a Kohn-Sham eigenstate for spin
channel $\sigma$ with band index $i$, crystal momentum $\bfk$ and
occupancy $f_{i\bfk}^{(\sigma)}$, and
$\hat{P}^{(I) m'}_m = \ket{\varphi^{(I)}_m}\bra{\varphi^{(I) m'}}$
is the Hubbard projection operator.
The contravariant dual vectors $\ket{\varphi^{(I) m}}$ are
related to the covariant projectors through the site-centered overlap
matrix $O^{\left( I \right)}_{m m' } = \braket{\varphi^{(I)}_m}{\varphi^{(I)}_{m'}}$
which is a metric on the correlated subspace
$\mathcal{C}^{(I)}$:
$\ket{\varphi^{(I) m}} = \ket{\varphi^{(I)}_{m'}} O^{(I) m' m}$;
$O^{(I) m' m''} O^{(I)}_{m'' m} = \delta^{m'}_{m}$. 

Our definition of the occupancy matrix differs to that of
Refs.~\onlinecite{PhysRevB.73.045110,*eschrig} and has the following
desirable properties: the expressions are tensorially correct; the
energy and resulting potential are rotationally invariant; the
resulting potential is Hermitian and localized to the correlated site;
and the trace of the occupancy matrix gives the occupancy of the
correlated 
site~\cite{forthcoming}.
The contravariant metric 
$O^{(I)m  m'}$ is calculated only as an inverse of the covariant
overlap matrix $O^{(I)}_{m m'}$, therefore, the duals of the Hubbard
projectors are also localized to the site.
As a result, and in contrast with previously proposed
approaches to DFT+$U$ models using non-orthogonal projectors, the
DFT+$U$ potential constructed from the tensorially consistent energy
for a given correlated site remains manifestly local to that site.
We note that for the special case of an orthogonal set of projectors
on each site, the projection operator is self-adjoint and the
above expressions reduce to the DFT+$U$ correction of
Ref.~\onlinecite{PhysRevB.71.035105,*PhysRevLett.97.103001}.

Any set of localized functions may, in principle, be used as Hubbard 
projectors with which to define the occupancy
matrix. Solutions of appropriate
orbital symmetry of the hydrogenic Schr\"{o}dinger equation,
such as atomic-like or linear muffin-tin orbitals,
 are a common 
 choice~\cite{PhysRevB.44.943, PhysRevB.58.1201, PhysRevB.57.1505}.
These are generally 
characterized by an effective charge $Z$
that determines their spatial diffuseness.
For a given value of $U$, results of DFT+$U$
calculations with different values chosen for $Z$
 will not necessarily yield the same ground-state
properties~\cite{PhysRevB.58.1201,PhysRevB.72.237102}.
Notwithstanding, hydrogenic orbitals may be
inappropriate in cases in which the orbitals of the correlated manifold
 differ significantly from atomic wavefunctions. 

In order to obtain accurate occupancies, a set of projectors is
required which adequately accounts for electronic hybridization
and which, if possible, is defined unambiguously for the system under
study. Wannier functions, in
particular maximally-localized Wannier functions
(MLWFs)~\cite{PhysRevB.56.12847},
 form just such an accurate 
minimal basis.
They have been used with good effect to 
augment DFT with localized 
many-body interactions~\cite{PhysRevB.74.125120}, 
and there is numerical evidence to suggest that MLWFs constitute the
projector set which maximizes the
$U$ parameter~\cite{PhysRevB.77.085122}.

We work with the single-particle
density-matrix, which is expressed in separable form~\cite{mcweeny}
$\rho(\mathbf{r},\mathbf{r'}) =
\sum_{\alpha\beta} \phi_{\alpha}(\mathbf{r}) K^{\alpha\beta}
\phi_{\beta}(\mathbf{r'})$
in terms of a localized basis
of NGWFs~\cite{PhysRevB.66.035119} $\{ \phi_{\alpha}(\mathbf{r}) \}$,
related to the Kohn-Sham eigenstates by a linear
transformation $\psi^{(\sigma)}_{n}(\mathbf{r}) =  
\sum_{\alpha} \phi_\alpha (\mathbf{r}) M^{(\sigma) \alpha}_{n}$. 
The density kernel $K^{\alpha\beta} = \langle \phi^{\alpha} \rvert
 \hat{\rho} \lvert \phi^{\beta} \rangle$ 
 is the representation of the single-particle density operator $\hat{\rho}$
 in terms of the contravariant duals $\{ \phi^{\alpha} (\mathbf{r})\}$
 of the NGWFs,
  which satisfy 
 $\langle \phi_{\alpha} \rvert \phi^{\beta} \rangle
  = \delta_{\alpha}^{\beta}$.
  The NGWFs are in turn expanded in terms of a 
  systematic basis of
Fourier-Lagrange, or psinc~\cite{mostofi-jcp03,*Baye1986},
functions.
The size of this basis is determined by an energy cutoff, 
akin to a plane-wave kinetic energy cutoff,
with respect to which calculations are converged.
 The DFT energy functional 
is iteratively minimized with respect
to both the density kernel and the NGWF expansion coefficients.
The minimization scheme in
the ONETEP linear-scaling
code is detailed in Refs.~\onlinecite{onetep1,*ChemPhysLett.422.345}.

These NGWFs, therefore, are a readily
accessible set of localized orbitals which are calculated with
linear-scaling computational cost.
Similarly to MLWFs, NGWF centres may be used to calculate
polarizabilities~\cite{forthcoming}. Thus, in this framework,
it is natural to use 
a localised subset of Wannier functions obtained at the
end of a ground-state calculation, with appropriate 
orbital character, as Hubbard projectors for
defining the DFT+$U$ occupancy matrix. NGWFs are adapted
to their chemical environment, reflecting the balance between the 
competing tendencies of electron itinerancy and localization in 
strongly correlated systems and, as a result, provide an accurate
representation of the occupancy of the correlated site.

We propose a projector self-consistent scheme whereby the Hubbard
projectors are determined self-consistently by iteratively solving for
the Kohn-Sham ground-state using the Hubbard projectors defined by
NGWFs from the DFT+$U$ ground-state energy calculation of the previous
iteration. In this way, the Hubbard projectors converge to those
that are optimally adapted for their own DFT+$U$ ground-state density.
This scheme, as we go on to show, rapidly and monotonically
converges to an unambiguously defined DFT+$U$ ground-state which, for
a given $U$ parameter, is of lowest energy. In other words, the
DFT+$U$ energy functional is additionally minimized with respect to
the set of localized NGWF projectors that are, at convergence, self-consistent
with the DFT+$U$ calculation from which they are determined.

We applied our method to iron porphyrin (FeP). Metalloporphyrin systems,
such as FeP, play an important role in biochemistry. 
The ability of metalloporphyrins to bind simple molecules is of
interest, particularly in the case of FeP which can have a greater
affinity for CO and NO than O$_2$, resulting in hindrance of
respiration.

We performed fully converged energy minimization on FeP, and its
complex with carbon monoxide, 
using the ONETEP code~\cite{onetep1}. 
We used spin-polarized DFT+$U$ within the generalized-gradient
(GGA)~\cite{PhysRevLett.77.3865} and
pseudopotential~\cite{PhysRevB.41.1227,*opium} approximations.
An equivalent
plane-wave kinetic energy cutoff of 1000~eV was used with a cubic
simulation cell of side-length 37~\AA.
The NGWFs were spatially restricted to atom-centered spheres of
radius 5.3~\AA \phantom{} 
and no density kernel truncation was
applied. Since the principal focus of this study was
the dependence of computed DFT+U ground-state properties
on variations in the Hubbard projectors for a given $U$ value, 
optimized PBE ($U=0$~eV) structures were used. 

\begin{figure}[t!]
\includegraphics*[width=8.6cm]
{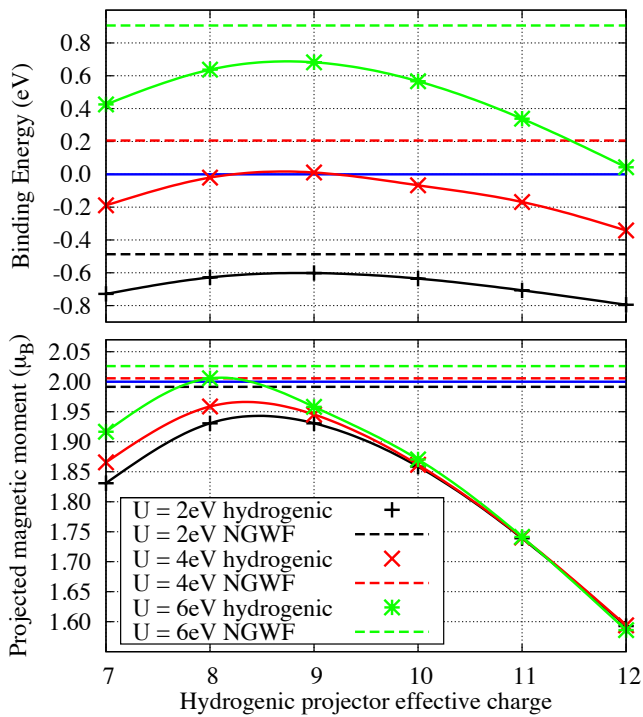}
\caption{
(Color online)
The interaction energy,
positive for an unbound ligand,
  of the CO and FeP moieties (top panel) and 
the magnetic dipole moment projected 
onto the correlated manifold of triplet-state FeP  
(bottom panel). 
  Both are plotted at various $U$ as a function of the
  effective charge $Z$ used to define the 
  hydrogenic projectors (solid lines), while 
  dashed lines show those quantities 
  calculated with self-consistent
  NGWF Hubbard projectors.
  Blue lines indicate the binding threshold (top)
  and the ideal projected moment (bottom).}
\label{Fig:FeXPo_magnetisation_vs_Z}
\end{figure}

Shown in
Fig.~\ref{Fig:FeXPo_magnetisation_vs_Z}, 
is the interaction energy between FeP and CO 
as an illustration that the binding affinity
 between moieties in DFT+$U$ can be strongly influenced
by the localization of the Hubbard projectors.
As can be seen, binding affinity is by no means
uniquely defined when 
hydrogenic projectors are used, although this
may be partly compensated by a projector-dependent 
first-principles~\cite{
PhysRevB.71.035105,*PhysRevLett.97.103001,
PhysRevB.58.1201} $U$ parameter.
At $U=6$~eV it varies from approximately 
0.04~eV to 0.69~eV over the
range of $Z$ considered; at $U=4$~eV the result is even qualitatively
ambiguous as a function of $Z$. 
Using self-consistent NGWF
projectors (dashed lines) generally results in 
energetically less favourable ligand binding, 
demonstrating that, for a given value of $U$, NGWF
projectors more effectively counteract the spurious tendency
of GGA functionals to over-bind ligands to FeP~\cite{scherlis}. 
Also shown in Fig.~\ref{Fig:FeXPo_magnetisation_vs_Z},
the projected magnetic dipole moment of FeP 
in its ground-state
varies strongly with the value of $Z$ 
chosen for hydrogenic projectors
(solid line), with only a narrow range
of $Z$ at $U=6$~eV giving values that are close to 
the expected $2.0\: \mu_{\rm B}$ for optimal projectors.
Moreover, a pathological inconsistency with experiment
emerges in that $U$ values of sufficient magnitude to achieve
the requisite moment (for some $Z$) bring us into the unphysical
regime where FeP+CO binding is disfavoured.
Conversely, the use of self-consistent NGWF 
projectors (dashed) 
results in a projected magnetic moment 
which lies within 
the physically reasonable range
and is rather insensitive to $U$.

\begin{figure}[t!]
\includegraphics*[width=8.5cm]
{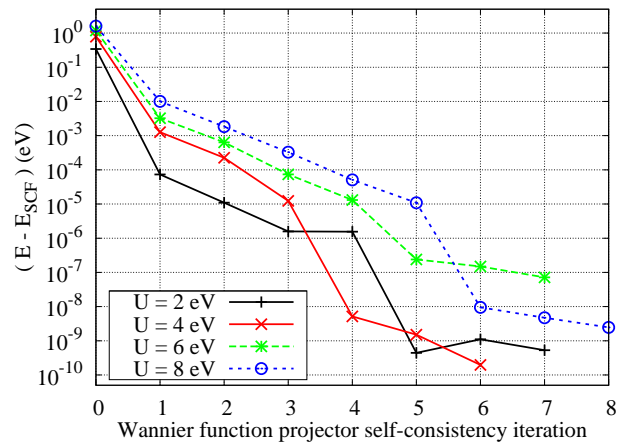}
\caption{
(Color online)
The difference in total energy $E$ and the total energy at
  projector self-consistency $E_{\rm SCF}$ as a function of the projector
  self-consistency iteration.
  The procedure is initialized (iteration 0) with a 
  set of hydrogenic $3d$
projectors to construct the correlated subspace, using the
Clementi-Raimondi~\cite{clementi} 
effective charge of $Z=11.17$ for
iron $3d$ orbitals. }
\label{Fig:FeCOPo_HUBBARDSCF_convergence}
\end{figure}

Fig.~\ref{Fig:FeCOPo_HUBBARDSCF_convergence} 
demonstrates the stable convergence of the Hubbard projector
self-consistency scheme for FeP+CO at different values of $U$. Each
data point represents an individual variational total-energy
minimization, wherein the Hubbard projectors are re-constructed 
from the optimized ground-state NGWFs from the previous iteration.  
The energy decreases rapidly as the projectors are refined,
converging within a small number of iterations. This
confirms our understanding that the Wannier are optimally adapted
for the hybridized character of the electronic orbitals, while
minimizing the energy. In this way, more spatially 
diffuse self-interaction corrections are introduced
than with purely atomic orbitals,
in a complimentary manner to such methods as 
DFT+$U$+$V$~\cite{0953-8984-22-5-055602} which 
allow more general interaction terms between sites.

\begin{figure}[t!]
\includegraphics*[width=8.5cm]
{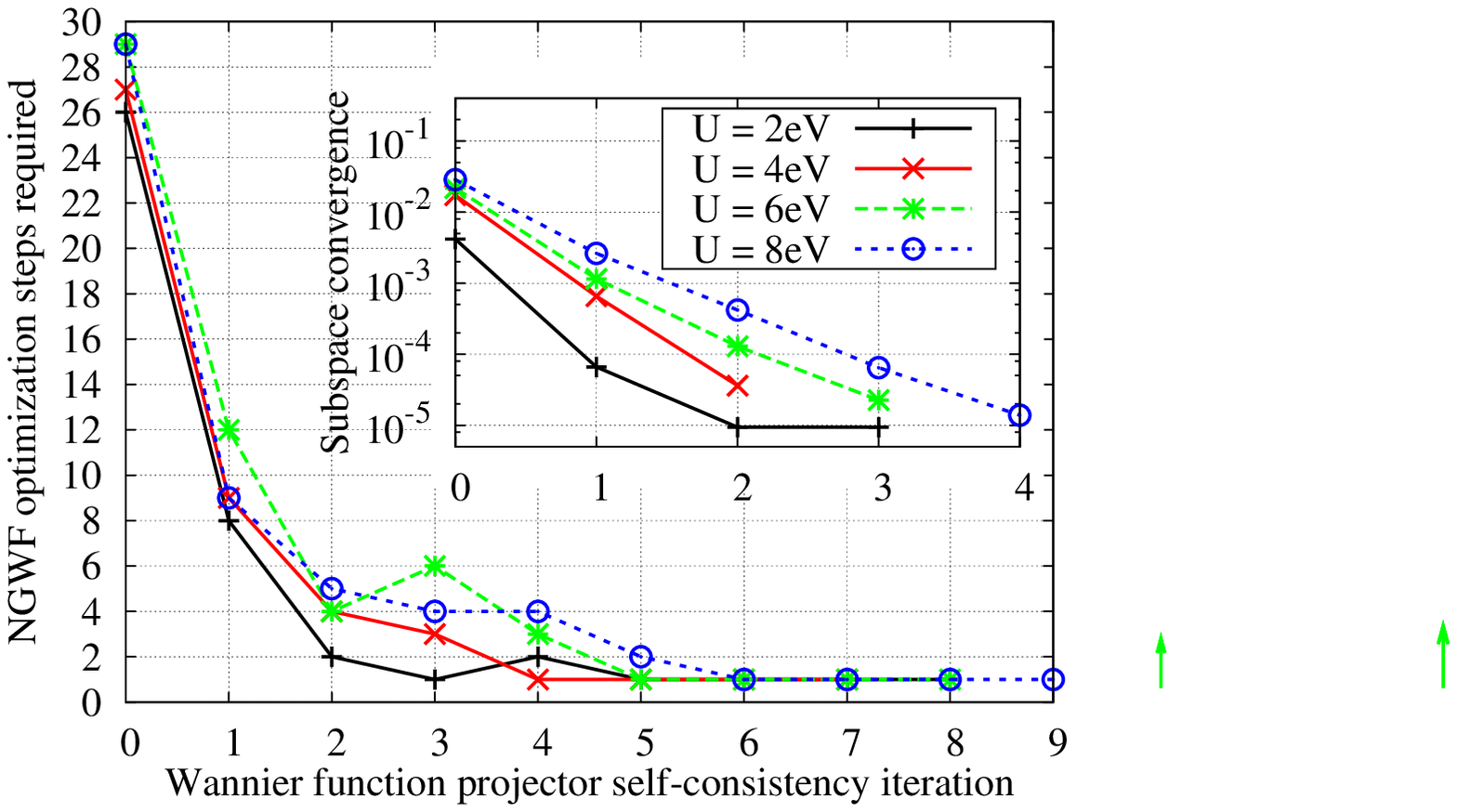}
\caption{
(Color online)
The number of NGWF optimization steps needed 
to converge the total energy for each projector self-consistency iteration for
  FeP+CO. Shown inset is the convergence of the correlated
  subspace, as quantified by its $3d$-orbital character.
   }
\label{Fig:FeCOPo_HUBBARDSCF_NGWFITER}
\end{figure}

Since we re-use the self-consistent density from the
previous projector iteration to initialise the following iteration,
much fewer NGWF optimization steps are required at each successive
projector update step. As demonstrated in  
Fig.~\ref{Fig:FeCOPo_HUBBARDSCF_NGWFITER}, this results in
an overall computational effort for achieving projector
self-consistency 
that is only a small overhead compared to the conventional approach.

In order to achieve meaningful insight into the $U$-dependence
of bond formation, 
it is necessary to allow for Hubbard projector update 
consistent with variations in the molecular geometry.
We stress that ionic force expressions are not 
complicated by the inclusion of  self-consistent 
Hubbard projectors, with no additional terms 
appearing over those in conventional DFT+$U$.

In conclusion, we have proposed and demonstrated a method within
DFT+$U$ for obtaining Hubbard projectors that are uniquely-defined, 
optimally adapted to their chemical environment, and consistent
with the DFT+$U$ ground-state density. Our
implementation may be
incorporated into any method that either solves directly for
localized Wannier-like states, or which computes such states
in a post-processing fashion. If combined self-consistently with 
approaches for calculating $U$ from 
first-principles~\cite{
PhysRevB.71.035105,*PhysRevLett.97.103001,
PhysRevB.58.1201}, this work opens
up the possibility of parameter-free DFT+$U$ calculations on large
systems.

\begin{acknowledgments}
This research was supported by EPSRC, 
RCUK and the National University of Ireland.
Calculations were performed on the Cambridge HPCS 
Darwin computer
under EPSRC grant EP/F032773/1.
\end{acknowledgments}


\begin{thebibliography}{38}%
\makeatletter
\providecommand \@ifxundefined [1]{%
 \@ifx{#1\undefined}
}%
\providecommand \@ifnum [1]{%
 \ifnum #1\expandafter \@firstoftwo
 \else \expandafter \@secondoftwo
 \fi
}%
\providecommand \@ifx [1]{%
 \ifx #1\expandafter \@firstoftwo
 \else \expandafter \@secondoftwo
 \fi
}%
\providecommand \natexlab [1]{#1}%
\providecommand \enquote  [1]{``#1''}%
\providecommand \bibnamefont  [1]{#1}%
\providecommand \bibfnamefont [1]{#1}%
\providecommand \citenamefont [1]{#1}%
\providecommand \href@noop [0]{\@secondoftwo}%
\providecommand \href [0]{\begingroup \@sanitize@url \@href}%
\providecommand \@href[1]{\@@startlink{#1}\@@href}%
\providecommand \@@href[1]{\endgroup#1\@@endlink}%
\providecommand \@sanitize@url [0]{\catcode `\\12\catcode `\$12\catcode
  `\&12\catcode `\#12\catcode `\^12\catcode `\_12\catcode `\%12\relax}%
\providecommand \@@startlink[1]{}%
\providecommand \@@endlink[0]{}%
\providecommand \url  [0]{\begingroup\@sanitize@url \@url }%
\providecommand \@url [1]{\endgroup\@href {#1}{\urlprefix }}%
\providecommand \urlprefix  [0]{URL }%
\providecommand \Eprint [0]{\href }%
\@ifxundefined \urlstyle {%
  \providecommand \doi  [0]{\begingroup \@sanitize@url \@doi}%
  \providecommand \@doi [1]{\endgroup \@@startlink {\doibase
  #1}doi:\discretionary {}{}{}#1\@@endlink }%
}{%
  \providecommand \doi  [0]{doi:\discretionary{}{}{}\begingroup
  \urlstyle{rm}\Url }%
}%
\providecommand \doibase [0]{http://dx.doi.org/}%
\providecommand \Doi [0]{\begingroup \@sanitize@url \@Doi }%
\providecommand \@Doi  [1]{\endgroup\@@startlink{\doibase#1}\@@Doi}%
\providecommand \@@Doi [1]{#1\@@endlink}%
\providecommand \selectlanguage [0]{\@gobble}%
\providecommand \bibinfo  [0]{\@secondoftwo}%
\providecommand \bibfield  [0]{\@secondoftwo}%
\providecommand \translation [1]{[#1]}%
\providecommand \BibitemOpen [0]{}%
\providecommand \bibitemStop [0]{}%
\providecommand \bibitemNoStop [0]{.\EOS\space}%
\providecommand \EOS [0]{\spacefactor3000\relax}%
\providecommand \BibitemShut  [1]{\csname bibitem#1\endcsname}%
\bibitem [{\citenamefont {Miller}\ and\ \citenamefont
  {Epstein}(1994)}]{angewchem.33.385}%
  \BibitemOpen
  \bibfield  {author} {\bibinfo {author} {\bibfnamefont {J.~S.}\ \bibnamefont
  {Miller}}\ and\ \bibinfo {author} {\bibfnamefont {A.~J.}\ \bibnamefont
  {Epstein}},\ }\href@noop {} {\bibfield  {journal} {\bibinfo  {journal}
  {Angew. Chem. Int. Ed. Engl.},\ }\textbf {\bibinfo {volume} {33}},\ \bibinfo
  {pages} {385} (\bibinfo {year} {1994})}\BibitemShut {NoStop}%
\bibitem [{\citenamefont {Steele}\ and\ \citenamefont
  {Heinzel}(2001)}]{nature.414.345}%
  \BibitemOpen
  \bibfield  {author} {\bibinfo {author} {\bibfnamefont {B.~C.~H.}\
  \bibnamefont {Steele}}\ and\ \bibinfo {author} {\bibfnamefont
  {A.}~\bibnamefont {Heinzel}},\ }\href@noop {} {\bibfield  {journal} {\bibinfo
   {journal} {Nature},\ }\textbf {\bibinfo {volume} {414}},\ \bibinfo {pages}
  {345} (\bibinfo {year} {2001})}\BibitemShut {NoStop}%
\bibitem [{\citenamefont {Holm}\ \emph {et~al.}(1996)\citenamefont {Holm},
  \citenamefont {Kennepohl},\ and\ \citenamefont {Solomon}}]{chemrev.96.2239}%
  \BibitemOpen
  \bibfield  {author} {\bibinfo {author} {\bibfnamefont {R.~H.}\ \bibnamefont
  {Holm}}, \bibinfo {author} {\bibfnamefont {P.}~\bibnamefont {Kennepohl}}, \
  and\ \bibinfo {author} {\bibfnamefont {E.~I.}\ \bibnamefont {Solomon}},\
  }\href@noop {} {\bibfield  {journal} {\bibinfo  {journal} {Chem. Rev.},\
  }\textbf {\bibinfo {volume} {96}},\ \bibinfo {pages} {2239} (\bibinfo {year}
  {1996})}\BibitemShut {NoStop}%
\bibitem [{\citenamefont {Hohenberg}\ and\ \citenamefont
  {Kohn}(1964)}]{PhysRev.136.B864}%
  \BibitemOpen
  \bibfield  {author} {\bibinfo {author} {\bibfnamefont {P.}~\bibnamefont
  {Hohenberg}}\ and\ \bibinfo {author} {\bibfnamefont {W.}~\bibnamefont
  {Kohn}},\ }\Doi {10.1103/PhysRev.136.B864} {\bibfield  {journal} {\bibinfo
  {journal} {Phys. Rev.},\ }\textbf {\bibinfo {volume} {136}},\ \bibinfo
  {pages} {B864} (\bibinfo {year} {1964})}\BibitemShut {NoStop}%
\bibitem [{\citenamefont {Kohn}\ and\ \citenamefont
  {Sham}(1965)}]{PhysRev.140.A1133}%
  \BibitemOpen
  \bibfield  {author} {\bibinfo {author} {\bibfnamefont {W.}~\bibnamefont
  {Kohn}}\ and\ \bibinfo {author} {\bibfnamefont {L.~J.}\ \bibnamefont
  {Sham}},\ }\Doi {10.1103/PhysRev.140.A1133} {\bibfield  {journal} {\bibinfo
  {journal} {Phys. Rev.},\ }\textbf {\bibinfo {volume} {140}},\ \bibinfo
  {pages} {A1133} (\bibinfo {year} {1965})}\BibitemShut {NoStop}%
\bibitem [{\citenamefont {Terakura}\ \emph {et~al.}(1984)\citenamefont
  {Terakura}, \citenamefont {Oguchi}, \citenamefont {Williams},\ and\
  \citenamefont {K\"ubler}}]{PhysRevB.30.4734}%
  \BibitemOpen
  \bibfield  {author} {\bibinfo {author} {\bibfnamefont {K.}~\bibnamefont
  {Terakura}}, \bibinfo {author} {\bibfnamefont {T.}~\bibnamefont {Oguchi}},
  \bibinfo {author} {\bibfnamefont {A.~R.}\ \bibnamefont {Williams}}, \ and\
  \bibinfo {author} {\bibfnamefont {J.}~\bibnamefont {K\"ubler}},\ }\Doi
  {10.1103/PhysRevB.30.4734} {\bibfield  {journal} {\bibinfo  {journal} {Phys.
  Rev. B},\ }\textbf {\bibinfo {volume} {30}},\ \bibinfo {pages} {4734}
  (\bibinfo {year} {1984})}\BibitemShut {NoStop}%
\bibitem [{\citenamefont {Dudarev}\ \emph {et~al.}(1998)\citenamefont
  {Dudarev}, \citenamefont {Botton}, \citenamefont {Savrasov}, \citenamefont
  {Humphreys},\ and\ \citenamefont {Sutton}}]{PhysRevB.57.1505}%
  \BibitemOpen
  \bibfield  {author} {\bibinfo {author} {\bibfnamefont {S.~L.}\ \bibnamefont
  {Dudarev}}, \bibinfo {author} {\bibfnamefont {G.~A.}\ \bibnamefont {Botton}},
  \bibinfo {author} {\bibfnamefont {S.~Y.}\ \bibnamefont {Savrasov}}, \bibinfo
  {author} {\bibfnamefont {C.~J.}\ \bibnamefont {Humphreys}}, \ and\ \bibinfo
  {author} {\bibfnamefont {A.~P.}\ \bibnamefont {Sutton}},\ }\Doi
  {10.1103/PhysRevB.57.1505} {\bibfield  {journal} {\bibinfo  {journal} {Phys.
  Rev. B},\ }\textbf {\bibinfo {volume} {57}},\ \bibinfo {pages} {1505}
  (\bibinfo {year} {1998})}\BibitemShut {NoStop}%
\bibitem [{\citenamefont {Perdew}\ \emph {et~al.}(1982)\citenamefont {Perdew},
  \citenamefont {Parr}, \citenamefont {Levy},\ and\ \citenamefont
  {Balduz}}]{PhysRevLett.49.1691}%
  \BibitemOpen
  \bibfield  {author} {\bibinfo {author} {\bibfnamefont {J.~P.}\ \bibnamefont
  {Perdew}}, \bibinfo {author} {\bibfnamefont {R.~G.}\ \bibnamefont {Parr}},
  \bibinfo {author} {\bibfnamefont {M.}~\bibnamefont {Levy}}, \ and\ \bibinfo
  {author} {\bibfnamefont {J.~L.}\ \bibnamefont {Balduz}},\ }\Doi
  {10.1103/PhysRevLett.49.1691} {\bibfield  {journal} {\bibinfo  {journal}
  {Phys. Rev. Lett.},\ }\textbf {\bibinfo {volume} {49}},\ \bibinfo {pages}
  {1691} (\bibinfo {year} {1982})}\BibitemShut {NoStop}%
\bibitem [{\citenamefont {Cococcioni}\ and\ \citenamefont
  {de~Gironcoli}(2005)}]{PhysRevB.71.035105}%
  \BibitemOpen
  \bibfield  {author} {\bibinfo {author} {\bibfnamefont {M.}~\bibnamefont
  {Cococcioni}}\ and\ \bibinfo {author} {\bibfnamefont {S.}~\bibnamefont
  {de~Gironcoli}},\ }\Doi {10.1103/PhysRevB.71.035105} {\bibfield  {journal}
  {\bibinfo  {journal} {Phys. Rev. B},\ }\textbf {\bibinfo {volume} {71}},\
  \bibinfo {pages} {035105} (\bibinfo {year} {2005})}\BibitemShut {NoStop}%
\bibitem [{\citenamefont {Kulik}\ \emph {et~al.}(2006)\citenamefont {Kulik},
  \citenamefont {Cococcioni}, \citenamefont {Scherlis},\ and\ \citenamefont
  {Marzari}}]{PhysRevLett.97.103001}%
  \BibitemOpen
  \bibfield  {author} {\bibinfo {author} {\bibfnamefont {H.~J.}\ \bibnamefont
  {Kulik}}, \bibinfo {author} {\bibfnamefont {M.}~\bibnamefont {Cococcioni}},
  \bibinfo {author} {\bibfnamefont {D.~A.}\ \bibnamefont {Scherlis}}, \ and\
  \bibinfo {author} {\bibfnamefont {N.}~\bibnamefont {Marzari}},\ }\Doi
  {10.1103/PhysRevLett.97.103001} {\bibfield  {journal} {\bibinfo  {journal}
  {Phys. Rev. Lett.},\ }\textbf {\bibinfo {volume} {97}},\ \bibinfo {pages}
  {103001} (\bibinfo {year} {2006})}\BibitemShut {NoStop}%
\bibitem [{\citenamefont {Cohen}\ \emph {et~al.}(2008)\citenamefont {Cohen},
  \citenamefont {Mori-Sanchez},\ and\ \citenamefont
  {Yang}}]{AronJ.Cohen08082008}%
  \BibitemOpen
  \bibfield  {author} {\bibinfo {author} {\bibfnamefont {A.~J.}\ \bibnamefont
  {Cohen}}, \bibinfo {author} {\bibfnamefont {P.}~\bibnamefont {Mori-Sanchez}},
  \ and\ \bibinfo {author} {\bibfnamefont {W.}~\bibnamefont {Yang}},\ }\Doi
  {10.1126/science.1158722} {\bibfield  {journal} {\bibinfo  {journal}
  {Science},\ }\textbf {\bibinfo {volume} {321}},\ \bibinfo {pages} {792}
  (\bibinfo {year} {2008})}\BibitemShut {NoStop}%
\bibitem [{\citenamefont {Svane}\ and\ \citenamefont
  {Gunnarsson}(1990)}]{PhysRevLett.65.1148}%
  \BibitemOpen
  \bibfield  {author} {\bibinfo {author} {\bibfnamefont {A.}~\bibnamefont
  {Svane}}\ and\ \bibinfo {author} {\bibfnamefont {O.}~\bibnamefont
  {Gunnarsson}},\ }\Doi {10.1103/PhysRevLett.65.1148} {\bibfield  {journal}
  {\bibinfo  {journal} {Phys. Rev. Lett.},\ }\textbf {\bibinfo {volume} {65}},\
  \bibinfo {pages} {1148} (\bibinfo {year} {1990})}\BibitemShut {NoStop}%
\bibitem [{\citenamefont {Anisimov}\ \emph {et~al.}(1991)\citenamefont
  {Anisimov}, \citenamefont {Zaanen},\ and\ \citenamefont
  {Andersen}}]{PhysRevB.44.943}%
  \BibitemOpen
  \bibfield  {author} {\bibinfo {author} {\bibfnamefont {V.~I.}\ \bibnamefont
  {Anisimov}}, \bibinfo {author} {\bibfnamefont {J.}~\bibnamefont {Zaanen}}, \
  and\ \bibinfo {author} {\bibfnamefont {O.~K.}\ \bibnamefont {Andersen}},\
  }\Doi {10.1103/PhysRevB.44.943} {\bibfield  {journal} {\bibinfo  {journal}
  {Phys. Rev. B},\ }\textbf {\bibinfo {volume} {44}},\ \bibinfo {pages} {943}
  (\bibinfo {year} {1991})}\BibitemShut {NoStop}%
\bibitem [{\citenamefont {Anisimov}\ \emph {et~al.}(1993)\citenamefont
  {Anisimov}, \citenamefont {Solovyev}, \citenamefont {Korotin}, \citenamefont
  {Czy\ifmmode~\dot{z}\else \.{z}\fi{}yk},\ and\ \citenamefont
  {Sawatzky}}]{PhysRevB.48.16929}%
  \BibitemOpen
  \bibfield  {author} {\bibinfo {author} {\bibfnamefont {V.~I.}\ \bibnamefont
  {Anisimov}}, \bibinfo {author} {\bibfnamefont {I.~V.}\ \bibnamefont
  {Solovyev}}, \bibinfo {author} {\bibfnamefont {M.~A.}\ \bibnamefont
  {Korotin}}, \bibinfo {author} {\bibfnamefont {M.~T.}\ \bibnamefont
  {Czy\ifmmode~\dot{z}\else \.{z}\fi{}yk}}, \ and\ \bibinfo {author}
  {\bibfnamefont {G.~A.}\ \bibnamefont {Sawatzky}},\ }\Doi
  {10.1103/PhysRevB.48.16929} {\bibfield  {journal} {\bibinfo  {journal} {Phys.
  Rev. B},\ }\textbf {\bibinfo {volume} {48}},\ \bibinfo {pages} {16929}
  (\bibinfo {year} {1993})}\BibitemShut {NoStop}%
\bibitem [{\citenamefont {Hubbard}(1963)}]{hubbard1}%
  \BibitemOpen
  \bibfield  {author} {\bibinfo {author} {\bibfnamefont {J.}~\bibnamefont
  {Hubbard}},\ }\href@noop {} {\bibfield  {journal} {\bibinfo  {journal} {Proc.
  R. Soc. London Ser. A},\ }\textbf {\bibinfo {volume} {276}},\ \bibinfo
  {pages} {238} (\bibinfo {year} {1963})}\BibitemShut {NoStop}%
\bibitem [{\citenamefont {Hubbard}(1964){\natexlab{a}}}]{hubbard2}%
  \BibitemOpen
  \bibfield  {author} {\bibinfo {author} {\bibfnamefont {J.}~\bibnamefont
  {Hubbard}},\ }\href@noop {} {\bibfield  {journal} {\bibinfo  {journal} {Proc.
  R. Soc. London Ser. A},\ }\textbf {\bibinfo {volume} {277}},\ \bibinfo
  {pages} {237} (\bibinfo {year} {1964}{\natexlab{a}})}\BibitemShut {NoStop}%
\bibitem [{\citenamefont {Hubbard}(1964){\natexlab{b}}}]{hubbard3}%
  \BibitemOpen
  \bibfield  {author} {\bibinfo {author} {\bibfnamefont {J.}~\bibnamefont
  {Hubbard}},\ }\href@noop {} {\bibfield  {journal} {\bibinfo  {journal} {Proc.
  R. Soc. London Ser. A},\ }\textbf {\bibinfo {volume} {281}},\ \bibinfo
  {pages} {401} (\bibinfo {year} {1964}{\natexlab{b}})}\BibitemShut {NoStop}%
\bibitem [{\citenamefont {Miyake}\ and\ \citenamefont
  {Aryasetiawan}(2008)}]{PhysRevB.77.085122}%
  \BibitemOpen
  \bibfield  {author} {\bibinfo {author} {\bibfnamefont {T.}~\bibnamefont
  {Miyake}}\ and\ \bibinfo {author} {\bibfnamefont {F.}~\bibnamefont
  {Aryasetiawan}},\ }\Doi {10.1103/PhysRevB.77.085122} {\bibfield  {journal}
  {\bibinfo  {journal} {Phys. Rev. B},\ }\textbf {\bibinfo {volume} {77}},\
  \bibinfo {pages} {085122} (\bibinfo {year} {2008})}\BibitemShut {NoStop}%
\bibitem [{\citenamefont {{A. A. Mostofi, J. R. Yates, Y.-S. Lee, I. Souza, D.
  Vanderbilt, N. Marzari}}(2008)}]{wannier90}%
  \BibitemOpen
  \bibfield  {author} {\bibinfo {author} {\bibnamefont {{A. A. Mostofi, J. R.
  Yates, Y.-S. Lee, I. Souza, D. Vanderbilt, N. Marzari}}},\ }\href@noop {}
  {\bibfield  {journal} {\bibinfo  {journal} {Comp. Phys. Comm.},\ }\textbf
  {\bibinfo {volume} {178}},\ \bibinfo {pages} {685} (\bibinfo {year}
  {2008})}\BibitemShut {NoStop}%
\bibitem [{\citenamefont {Pickett}\ \emph {et~al.}(1998)\citenamefont
  {Pickett}, \citenamefont {Erwin},\ and\ \citenamefont
  {Ethridge}}]{PhysRevB.58.1201}%
  \BibitemOpen
  \bibfield  {author} {\bibinfo {author} {\bibfnamefont {W.~E.}\ \bibnamefont
  {Pickett}}, \bibinfo {author} {\bibfnamefont {S.~C.}\ \bibnamefont {Erwin}},
  \ and\ \bibinfo {author} {\bibfnamefont {E.~C.}\ \bibnamefont {Ethridge}},\
  }\Doi {10.1103/PhysRevB.58.1201} {\bibfield  {journal} {\bibinfo  {journal}
  {Phys. Rev. B},\ }\textbf {\bibinfo {volume} {58}},\ \bibinfo {pages} {1201}
  (\bibinfo {year} {1998})}\BibitemShut {NoStop}%
\bibitem [{\citenamefont {Han}\ \emph {et~al.}(2006)\citenamefont {Han},
  \citenamefont {Ozaki},\ and\ \citenamefont {Yu}}]{PhysRevB.73.045110}%
  \BibitemOpen
  \bibfield  {author} {\bibinfo {author} {\bibfnamefont {M.~J.}\ \bibnamefont
  {Han}}, \bibinfo {author} {\bibfnamefont {T.}~\bibnamefont {Ozaki}}, \ and\
  \bibinfo {author} {\bibfnamefont {J.}~\bibnamefont {Yu}},\ }\Doi
  {10.1103/PhysRevB.73.045110} {\bibfield  {journal} {\bibinfo  {journal}
  {Phys. Rev. B},\ }\textbf {\bibinfo {volume} {73}},\ \bibinfo {pages}
  {045110} (\bibinfo {year} {2006})}\BibitemShut {NoStop}%
\bibitem [{\citenamefont {H.~Eschrig}\ and\ \citenamefont
  {Chaplygin}(2003)}]{eschrig}%
  \BibitemOpen
  \bibfield  {author} {\bibinfo {author} {\bibfnamefont {K.~K.}\ \bibnamefont
  {H.~Eschrig}}\ and\ \bibinfo {author} {\bibfnamefont {I.}~\bibnamefont
  {Chaplygin}},\ }\href@noop {} {\bibfield  {journal} {\bibinfo  {journal}
  {Journal of Solid State Chemistry},\ }\textbf {\bibinfo {volume} {176}},\
  \bibinfo {pages} {482} (\bibinfo {year} {2003})}\BibitemShut {NoStop}%
\bibitem [{for()}]{forthcoming}%
  \BibitemOpen
  \href@noop {} {}\bibinfo {note} {{Further details to appear in a forthcoming
  paper.}}\BibitemShut {Stop}%
\bibitem [{\citenamefont {Fabris}\ \emph {et~al.}(2005)\citenamefont {Fabris},
  \citenamefont {de~Gironcoli}, \citenamefont {Baroni}, \citenamefont
  {Vicario},\ and\ \citenamefont {Balducci}}]{PhysRevB.72.237102}%
  \BibitemOpen
  \bibfield  {author} {\bibinfo {author} {\bibfnamefont {S.}~\bibnamefont
  {Fabris}}, \bibinfo {author} {\bibfnamefont {S.}~\bibnamefont
  {de~Gironcoli}}, \bibinfo {author} {\bibfnamefont {S.}~\bibnamefont
  {Baroni}}, \bibinfo {author} {\bibfnamefont {G.}~\bibnamefont {Vicario}}, \
  and\ \bibinfo {author} {\bibfnamefont {G.}~\bibnamefont {Balducci}},\ }\Doi
  {10.1103/PhysRevB.72.237102} {\bibfield  {journal} {\bibinfo  {journal}
  {Phys. Rev. B},\ }\textbf {\bibinfo {volume} {72}},\ \bibinfo {pages}
  {237102} (\bibinfo {year} {2005})}\BibitemShut {NoStop}%
\bibitem [{\citenamefont {Marzari}\ and\ \citenamefont
  {Vanderbilt}(1997)}]{PhysRevB.56.12847}%
  \BibitemOpen
  \bibfield  {author} {\bibinfo {author} {\bibfnamefont {N.}~\bibnamefont
  {Marzari}}\ and\ \bibinfo {author} {\bibfnamefont {D.}~\bibnamefont
  {Vanderbilt}},\ }\Doi {10.1103/PhysRevB.56.12847} {\bibfield  {journal}
  {\bibinfo  {journal} {Phys. Rev. B},\ }\textbf {\bibinfo {volume} {56}},\
  \bibinfo {pages} {12847} (\bibinfo {year} {1997})}\BibitemShut {NoStop}%
\bibitem [{\citenamefont {Lechermann}\ \emph {et~al.}(2006)\citenamefont
  {Lechermann}, \citenamefont {Georges}, \citenamefont {Poteryaev},
  \citenamefont {Biermann}, \citenamefont {Posternak}, \citenamefont
  {Yamasaki},\ and\ \citenamefont {Andersen}}]{PhysRevB.74.125120}%
  \BibitemOpen
  \bibfield  {author} {\bibinfo {author} {\bibfnamefont {F.}~\bibnamefont
  {Lechermann}}, \bibinfo {author} {\bibfnamefont {A.}~\bibnamefont {Georges}},
  \bibinfo {author} {\bibfnamefont {A.}~\bibnamefont {Poteryaev}}, \bibinfo
  {author} {\bibfnamefont {S.}~\bibnamefont {Biermann}}, \bibinfo {author}
  {\bibfnamefont {M.}~\bibnamefont {Posternak}}, \bibinfo {author}
  {\bibfnamefont {A.}~\bibnamefont {Yamasaki}}, \ and\ \bibinfo {author}
  {\bibfnamefont {O.~K.}\ \bibnamefont {Andersen}},\ }\Doi
  {10.1103/PhysRevB.74.125120} {\bibfield  {journal} {\bibinfo  {journal}
  {Phys. Rev. B},\ }\textbf {\bibinfo {volume} {74}},\ \bibinfo {pages}
  {125120} (\bibinfo {year} {2006})}\BibitemShut {NoStop}%
\bibitem [{\citenamefont {McWeeny}(1960)}]{mcweeny}%
  \BibitemOpen
  \bibfield  {author} {\bibinfo {author} {\bibfnamefont {R.}~\bibnamefont
  {McWeeny}},\ }\href@noop {} {\bibfield  {journal} {\bibinfo  {journal} {Rev.
  Mod. Phys.},\ }\textbf {\bibinfo {volume} {32}},\ \bibinfo {pages} {335}
  (\bibinfo {year} {1960})}\BibitemShut {NoStop}%
\bibitem [{\citenamefont {Skylaris}\ \emph {et~al.}(2002)\citenamefont
  {Skylaris}, \citenamefont {Mostofi}, \citenamefont {Haynes}, \citenamefont
  {Di\'eguez},\ and\ \citenamefont {Payne}}]{PhysRevB.66.035119}%
  \BibitemOpen
  \bibfield  {author} {\bibinfo {author} {\bibfnamefont {C.-K.}\ \bibnamefont
  {Skylaris}}, \bibinfo {author} {\bibfnamefont {A.~A.}\ \bibnamefont
  {Mostofi}}, \bibinfo {author} {\bibfnamefont {P.~D.}\ \bibnamefont {Haynes}},
  \bibinfo {author} {\bibfnamefont {O.}~\bibnamefont {Di\'eguez}}, \ and\
  \bibinfo {author} {\bibfnamefont {M.~C.}\ \bibnamefont {Payne}},\ }\Doi
  {10.1103/PhysRevB.66.035119} {\bibfield  {journal} {\bibinfo  {journal}
  {Phys. Rev. B},\ }\textbf {\bibinfo {volume} {66}},\ \bibinfo {pages}
  {035119} (\bibinfo {year} {2002})}\BibitemShut {NoStop}%
\bibitem [{\citenamefont {{A. A. Mostofi, P. D. Haynes, C.-K. Skylaris and M.
  C. Payne}}(2003)}]{mostofi-jcp03}%
  \BibitemOpen
  \bibfield  {author} {\bibinfo {author} {\bibnamefont {{A. A. Mostofi, P. D.
  Haynes, C.-K. Skylaris and M. C. Payne}}},\ }\href@noop {} {\bibfield
  {journal} {\bibinfo  {journal} {J. Chem. Phys.},\ }\textbf {\bibinfo {volume}
  {119}},\ \bibinfo {pages} {8842} (\bibinfo {year} {2003})}\BibitemShut
  {NoStop}%
\bibitem [{\citenamefont {Baye}\ and\ \citenamefont {Heenen}(1986)}]{Baye1986}%
  \BibitemOpen
  \bibfield  {author} {\bibinfo {author} {\bibfnamefont {D.}~\bibnamefont
  {Baye}}\ and\ \bibinfo {author} {\bibfnamefont {P.-H.}\ \bibnamefont
  {Heenen}},\ }\href@noop {} {\bibfield  {journal} {\bibinfo  {journal} {J.
  Phys. A: Math. Gen.},\ }\textbf {\bibinfo {volume} {19}},\ \bibinfo {pages}
  {2041} (\bibinfo {year} {1986})}\BibitemShut {NoStop}%
\bibitem [{\citenamefont {{C.-K. Skylaris, P. D. Haynes, A. A. Mostofi and M.
  C. Payne}}(2005)}]{onetep1}%
  \BibitemOpen
  \bibfield  {author} {\bibinfo {author} {\bibnamefont {{C.-K. Skylaris, P. D.
  Haynes, A. A. Mostofi and M. C. Payne}}},\ }\href@noop {} {\bibfield
  {journal} {\bibinfo  {journal} {J. Chem. Phys.},\ }\textbf {\bibinfo {volume}
  {122}},\ \bibinfo {pages} {084119} (\bibinfo {year} {2005})}\BibitemShut
  {NoStop}%
\bibitem [{\citenamefont {Haynes}\ \emph {et~al.}(2006)\citenamefont {Haynes},
  \citenamefont {Skylaris}, \citenamefont {Mostofi},\ and\ \citenamefont
  {Payne}}]{ChemPhysLett.422.345}%
  \BibitemOpen
  \bibfield  {author} {\bibinfo {author} {\bibfnamefont {P.~D.}\ \bibnamefont
  {Haynes}}, \bibinfo {author} {\bibfnamefont {C.-K.}\ \bibnamefont
  {Skylaris}}, \bibinfo {author} {\bibfnamefont {A.~A.}\ \bibnamefont
  {Mostofi}}, \ and\ \bibinfo {author} {\bibfnamefont {M.~C.}\ \bibnamefont
  {Payne}},\ }\href@noop {} {\bibfield  {journal} {\bibinfo  {journal} {Chem.
  Phys. Lett.},\ }\textbf {\bibinfo {volume} {422}},\ \bibinfo {pages} {345}
  (\bibinfo {year} {2006})}\BibitemShut {NoStop}%
\bibitem [{\citenamefont {Perdew}\ \emph {et~al.}(1996)\citenamefont {Perdew},
  \citenamefont {Burke},\ and\ \citenamefont
  {Ernzerhof}}]{PhysRevLett.77.3865}%
  \BibitemOpen
  \bibfield  {author} {\bibinfo {author} {\bibfnamefont {J.~P.}\ \bibnamefont
  {Perdew}}, \bibinfo {author} {\bibfnamefont {K.}~\bibnamefont {Burke}}, \
  and\ \bibinfo {author} {\bibfnamefont {M.}~\bibnamefont {Ernzerhof}},\ }\Doi
  {10.1103/PhysRevLett.77.3865} {\bibfield  {journal} {\bibinfo  {journal}
  {Phys. Rev. Lett.},\ }\textbf {\bibinfo {volume} {77}},\ \bibinfo {pages}
  {3865} (\bibinfo {year} {1996})}\BibitemShut {NoStop}%
\bibitem [{\citenamefont {Rappe}\ \emph {et~al.}(1990)\citenamefont {Rappe},
  \citenamefont {Rabe}, \citenamefont {Kaxiras},\ and\ \citenamefont
  {Joannopoulos}}]{PhysRevB.41.1227}%
  \BibitemOpen
  \bibfield  {author} {\bibinfo {author} {\bibfnamefont {A.~M.}\ \bibnamefont
  {Rappe}}, \bibinfo {author} {\bibfnamefont {K.~M.}\ \bibnamefont {Rabe}},
  \bibinfo {author} {\bibfnamefont {E.}~\bibnamefont {Kaxiras}}, \ and\
  \bibinfo {author} {\bibfnamefont {J.~D.}\ \bibnamefont {Joannopoulos}},\
  }\Doi {10.1103/PhysRevB.41.1227} {\bibfield  {journal} {\bibinfo  {journal}
  {Phys. Rev. B},\ }\textbf {\bibinfo {volume} {41}},\ \bibinfo {pages} {1227}
  (\bibinfo {year} {1990})}\BibitemShut {NoStop}%
\bibitem [{opi()}]{opium}%
  \BibitemOpen
  \href@noop {} {}\bibinfo {note} {RRKJ Pseudopotentials were generated using
  the Opium code, {\tt http://opium.sourceforge.net}, using the GGA input
  parameters available therein, albeit with a scalar-relativistic correction
  for all species and, for iron, a non-linear core correction of
  Fuchs-Scheffler characteristic radius 1.3a.u. and core-radius of
  2.0a.u.}\BibitemShut {Stop}%
\bibitem [{\citenamefont {{D. A. Scherlis, M. Cococcioni, P. Sit, N.
  Marzari}}(2007)}]{scherlis}%
  \BibitemOpen
  \bibfield  {author} {\bibinfo {author} {\bibnamefont {{D. A. Scherlis, M.
  Cococcioni, P. Sit, N. Marzari}}},\ }\href@noop {} {\bibfield  {journal}
  {\bibinfo  {journal} {J. Phys. Chem. B},\ }\textbf {\bibinfo {volume}
  {111}},\ \bibinfo {pages} {7384} (\bibinfo {year} {2007})}\BibitemShut
  {NoStop}%
\bibitem [{\citenamefont {Clementi}\ and\ \citenamefont
  {Raimondi}(1963)}]{clementi}%
  \BibitemOpen
  \bibfield  {author} {\bibinfo {author} {\bibfnamefont {E.}~\bibnamefont
  {Clementi}}\ and\ \bibinfo {author} {\bibfnamefont {D.~L.}\ \bibnamefont
  {Raimondi}},\ }\href@noop {} {\bibfield  {journal} {\bibinfo  {journal} {J.
  Chem. Phys.},\ }\textbf {\bibinfo {volume} {38}},\ \bibinfo {pages} {2686}
  (\bibinfo {year} {1963})}\BibitemShut {NoStop}%
\bibitem [{\citenamefont {\mbox{Campo} Jr}\ and\ \citenamefont
  {Cococcioni}(2010)}]{0953-8984-22-5-055602}%
  \BibitemOpen
  \bibfield  {author} {\bibinfo {author} {\bibfnamefont {V.~L.}\ \bibnamefont
  {\mbox{Campo} Jr}}\ and\ \bibinfo {author} {\bibfnamefont {M.}~\bibnamefont
  {Cococcioni}},\ }\href {http://stacks.iop.org/0953-8984/22/i=5/a=055602}
  {\bibfield  {journal} {\bibinfo  {journal} {Journal of Physics: Condensed
  Matter},\ }\textbf {\bibinfo {volume} {22}},\ \bibinfo {pages} {055602}
  (\bibinfo {year} {2010})}\BibitemShut {NoStop}%
\end{thebibliography}
\end{document}